\documentclass[12pt]{article}

\usepackage{amsmath,amssymb}
\usepackage{lmodern}
\usepackage{iftex}
\ifPDFTeX
  \usepackage[T1]{fontenc}
  \usepackage[utf8]{inputenc}
  \usepackage{textcomp} 
\else 
  \usepackage{unicode-math}
  \defaultfontfeatures{Scale=MatchLowercase}
  \defaultfontfeatures[\rmfamily]{Ligatures=TeX,Scale=1}
\fi
\usepackage{xcolor}
\ifx\paragraph\undefined\else
  \let\oldparagraph\paragraph
  \renewcommand{\paragraph}[1]{\oldparagraph{#1}\mbox{}}
\fi
\ifx\subparagraph\undefined\else
  \let\oldsubparagraph\subparagraph
  \renewcommand{\subparagraph}[1]{\oldsubparagraph{#1}\mbox{}}
\fi

\usepackage{longtable,booktabs,array}
\usepackage{calc} 
\usepackage{etoolbox}
\makeatletter
\patchcmd\longtable{\par}{\if@noskipsec\mbox{}\fi\par}{}{}
\makeatother
\IfFileExists{footnotehyper.sty}{\usepackage{footnotehyper}}{\usepackage{footnote}}
\makesavenoteenv{longtable}
\usepackage{graphicx}
\makeatletter
\def\maxwidth{\ifdim\Gin@nat@width>\linewidth\linewidth\else\Gin@nat@width\fi}
\def\maxheight{\ifdim\Gin@nat@height>\textheight\textheight\else\Gin@nat@height\fi}
\makeatother
\setkeys{Gin}{width=\maxwidth,height=\maxheight,keepaspectratio}
\makeatletter
\def\fps@figure{htbp}
\makeatother
\newlength{\cslhangindent}
\newlength{\csllabelwidth}
\newlength{\cslentryspacingunit} 
 {
  \setlength{\parindent}{0pt}
  \ifodd #1
  \let\oldpar\par
  \def\par{\hangindent=\cslhangindent\oldpar}
  \fi
  \setlength{\parskip}{#2\cslentryspacingunit}
 }%
 {}
\usepackage{calc}

\usepackage[noblocks]{authblk}

\usepackage{amsfonts,amssymb,amsbsy,latexsym,amsmath,tabulary,graphicx,times,caption,fancyhdr}
\usepackage{url,multirow,morefloats,floatflt,cancel,tfrupee}
\makeatletter

\usepackage{setspace}
\usepackage{lineno}

\AtBeginDocument{\@ifpackageloaded{textcomp}{}{\usepackage{textcomp}}}
\makeatother
\usepackage{colortbl}
\usepackage{pifont}
\usepackage{amsmath}

\usepackage[nointegrals]{wasysym}
\usepackage{array}
\usepackage{tabularx,ragged2e,rotating}
 \usepackage{booktabs}
 \usepackage{pdflscape}
\usepackage{mathtools}
\usepackage{setspace}
\renewcommand{\abstractname}{Abstract}
\usepackage{bm}
\usepackage{rotating}
\usepackage{hyperref}
\hypersetup{colorlinks,allcolors=black}
\usepackage{longtable}
\usepackage[nointegrals]{wasysym}
\usepackage{svg}
\usepackage{amsmath}
\usepackage{booktabs}
\usepackage{notoccite}
\usepackage[tableposition=bottom]{caption} 
\usepackage{rotating}
\usepackage{subfig}
\usepackage[a4paper, paperwidth=8.5in,
   paperheight=11in,
   top=1in,
   bottom=1in,
   left=1in,
   right=1in]{geometry}
\usepackage[sort,compress]{natbib}

\title{The Point Process Framework for Integrated Modelling of Biodiversity Data}

\author[1,2,*]{Kwaku Peprah Adjei}
\author[1,2]{Philip Mostert}
\author[3]{Jorge Sicacha Parada}
\author[1]{Emma Skarstein}
\author[1,2]{Robert B. O'Hara}

\affil[1]{Department of Mathematical Sciences, Norwegian University of Science and Technology, Norway}
\affil[2]{Center for Biodiversity Dynamics, Norwegian University of Science and Technology, Norway}
\affil[3]{Norwegian Computing Center, Oslo, Norway}
\affil[*]{Corresponding author: kadjeipeprah94@gmail.com}

\date{}
\begin{document}
\maketitle



\begin{abstract}
The quantity and types of biodiversity data being collected have increased in recent years. If we are to model and monitor biodiversity effectively, we need to respect how different data sets were collected, and effectively integrate these data types together. The framework that has emerged to do this is based on a point process formulation, with individuals as points and their distribution as a realisation of a random field. We describe this formulation and how the process model for the actual distribution is linked to the data that is collected through observation models. The observation models describe the data collection process, and its uncertainties and biases. We provide an example of using these methods to model species of Norwegian freshwater fish, which shows how integrated models can be adapted to the data we can collect. We summarise the modelling issues that arise and the approaches that could be taken to solve them.

\def\keywordstitle{Keywords}

\end{abstract}


\section{Introduction}

In recent years, there has been an influx of biodiversity data collected and hosted in openly accessible databases \citep{anderson2018biodiversity, farley2018situating, mandeville2021open}. These biodiversity data are collected from different sources: monitoring programs that employ different sampling protocols, museum collections and less structured citizen science collections. This biodiversity data is vital for predicting species distribution and making inferences about their ecology and conservation. The mapping process is, however, impeded by inherent problems in the way the data have been collected, for example, due to misidentification and errors in the data
collection process \citep{kosmala2016assessing, aceves2017accuracy, chambert2015modeling, royle2006generalized}, differences in the scale and resolution of the data
\cite[such as data collected at different spatial resolutions][]{boakes2010distorted, donaldson2016taxonomic, gonzalez2016estimating} and availability of the data in various formats
\citep{bowker2000mapping}. Despite these challenges, integrating data from different sources has been a growing field in ecology \citep{isaac2020data}.

Species distribution models (SDMs) are now widely used by ecologists and
conservation biologists to help us understand where in space and time
species presence can be expected \citep{higgins2012niche}. SDMs, growing
in their popularity, have however come under some scrutiny from various
perspectives (for example, see \citet{a2022species} and
\citet{robinson2017systematic} for a review of some of these
challenges). The problem of interest in this study is the challenge of
fitting SDMs without respecting the data type and data collection
protocol. For instance, most SDMs fitted with MaxEnt
\citep{Maxent1, Maxent2} and Biomod \citep{thuiller2003biomod} assume
that the data is generated from a `presence-only' process, i.e.~it
consists of reports of observations of the species of interest and no
information about possible absences, regardless of the data collection
process.

In practice, however, the available data are heterogeneous. For example,
presence-only observations, formal repeated surveys, atlas data and
expert range maps may all be available. Thus, we need methods to effectively combine the different data types in a single analysis. Advances along these lines have been made by \citet{Dorazio2014} and \citet{Fithian2015}, who both developed a similar framework to combine
presence-only data with occupancy data from repeated surveys. Along
similar lines, \citet{pagel2014} use presence-absence data to supplement
data on abundance to improve the modelling of range dynamics.
Presence-only data and range maps have also been combined by
\citet{Merow2017}.

Additionally, each of the available data types has their own quirks (observation errors) 
 that should be added to the integrated model individually. Failure to properly account for these observation errors results in biased inferences about the actual distribution of interest  \citep{simmonds2020more}. The sources of observation errors include uneven sampling effort
\citep{chakraborty2011point,sicacha2021spatial, ahmad2021integrated, simmonds2020more},
imperfect detection \citep{kery2004monitoring, mackenzie2002estimating}
and misclassification
\citep{wright2020modelling, chambert2015modeling, miller2011improving}.
By providing separate sub-models for each available data in the integrated model, information is shared across the data to produce identifiable model parameter estimates whilst accounting for the observation errors in each data 
\citep{fletcher2016integrated, Dorazio2014, guillera2017dealing, isaac2020data}. For example, data collected with preferential sampling (i.e. sampling more where the species is present) and presence-only data can be integrated with presence-absence
data, camera trap and telemetry data to account for the sampling bias
\citep{adde2021integrated, koshkina2017integrated, fletcher2016integrated, simmonds2020more}.
Also, repeated counts from multiple visits to a site and double-observer
or site occupancy protocol data can be used to account for imperfect
detection \citep{Dorazio2014, kery2004monitoring, case2017integrating}.
To account for misclassification, data with information on the
verification process, either from Machine learning algorithms or experts
can be used \citep{spiers2022estimating, wright2020modelling}.

Fitting a distribution model with different data types requires a single
model for the actual distribution, which can then map onto the different
observation models for each data type. A common approach among those
developing the models is to use a point process model for the actual
distribution
\citep[e.g.][]{Aarts2012, Fithian2012, Renner2013, Dorazio2014, Fithian2015},
reviewed by \citet{Renneretal2015}. This is a model in continuous space,
which has several advantages: (i) it removes the need to discretise the
disparate data into grid cells, with the accompanying loss of spatial
accuracy, (ii) a problem with presence-only models was the choice of
``pseudo-absences'', but with the point process formulation, their roles
are clarified as quadrature points in a numerical integration
\citep[see below]{Renner2013}.

The purpose of this paper is to present the statistical formulation of
data integration using the point process framework. We first present a
brief review of the data integration literature that use the point
process framework and collate the data types usually combined. Then, we
present the process and observation models using the point process
framework and present an example of an SDM fitted to data on freshwater
fish in Norwegian lakes.

\section{Brief review of literature}\label{brief-review-of-literature}

Before we describe the model framework, we briefly present a literature
review of integrated models that are fitted using the point process
framework and the various observation models used in these studies.

A conceptual workflow that integrates disparate datasets together to
combine the individual strengths of each in an SDM was communicated by
\citet{jetz2012integrating}. They believed that doing so was beneficial
to reduce geographic and environmental biases inherent in
single dataset types, as well as improving quality control and
cross-validation among the datasets. Since this call, a multitude of
techniques to integrate diverse data in a model-based framework have
been proposed through statistical and ecological journals.
\citet{fletcher2019practical} and \citet{miller2019recent} provided
introductions detailing some of the methods to combine data: an informed
prior approach \citep{marcantonio2016first, talluto2016cross}, an
auxiliary data approach
\citep{merow2016improving, regos2016predicting, huberman2020advances},
an ensemble model approach
\citep{douma2012towards, case2017integrating}, a correlation model
approach \citep{pacifici2017integrating} or through mere data pooling.

However, the method clearly predominating in literature is the one based
on the Poisson point process framework. The first paper to consider the
point process model approach for integrated distribution models was
\citep{Dorazio2014}, who used it to combine presence-only data with
planned survey data to account for sampling biases in the former. Since then,
numerous other papers have implemented similar methods using a range of
different data types. Table \ref{tab:ISDM_table references} gives a list
of some of these papers along with the data types integrated.

\begin{table}[ht]
\setlength\tabcolsep{1pt}
\begin{tabular}{|p{1.2in}|p{4.5in}|}
\toprule
Observation models & Citations \\ \midrule
Presence absence   &  \citet{Dorazio2014}, \citet{Fithian2015}, \citet{fletcher2016integrated}, \citet{koshkina2017integrated}, \citet{schank2017using}, \citet{pacifici2017integrating}, \citet{fletcher2019practical}, \citet{miller2019recent}, \citet{gelfand2019preferential}, \citet{peel2019reliable}, \citet{duncan2020land}, \citet{isaac2020data}, \citet{simmonds2020more}, \citet{chevalier2021data}, \citet{adde2021integrated}, \cite{bu2021not}, \citet{watson2021estimating}, \citet{gilbert2021integrating}, \citet{ahmad2021integrated}, \citet{fidino2022integrated}, \citet{morera2023bayesian}, \citet{grattarola2022integrating}, \citet{mostert2022intsdm}, \citet{grabow2022data}\\ 
Range maps         &  \citet{Merow2017} \\
Abundance             & \citet{giraud2016capitalizing}, \citet{makinen2018hierarchical}\\ 
Distance sampling  &  \citet{martino2021integration}, \citet{farr2021integrating}, \citet{pace2022seasonal}\\
    Other & \citet{bowler2019integrating} (Abundance and Presence absence), \citet{zulian2021integrating} (Presence absence only), \citet{rufener2021bridging} (Abundance only), \citet{cunningham2021quantifying} (Counts and Density), \citet{sultaire2022spatial} (Presence absence only), \citet{lauret2022integrated} (Distance sampling and Presence absence), \citet{cunningham2022dynamics} (Abundance only)\\\bottomrule
\end{tabular}
\caption{Observation models used in conjunction with presence-only data found across citations implementing the point-process framework for data integration.}
\label{tab:ISDM_table references}
\end{table}

It is evident through this short review that the data types mostly integrated
together are presence-only, presence-absence, range maps, abundance and
distance sampling occurrence records. The combination of the disparate
datasets are either with different data types (for example, abundance
and presence-only data) or the same data type but from different
sampling protocols. Therefore, we proceed to describe the point process
framework for the process model and the various observation process
models for the data types available.

\section{Point Process Framework}\label{point-process-framework}
Because we have to deal with both the ecological and observation
processes, it is natural to use a state-space formulation to separate the model into process and observation models. We define the
unobserved distribution as a spatial field, \(\lambda(\mathbf{s},t)\) in
a space \(\mathbf{s} \in \mathbb{R}^2\) and time \(t\) (here we will
consider time to be discrete).

For each of the $M$ datasets we want to use in our model, we need a data-generating model that links the
observations to the underlying state. That is, if dataset \(d\) is
\(Y_d\), we want \(Pr(Y_d | \lambda(\mathbf{s},t), \theta_d)\), where
\(\theta_d\) are parameters of the data generating model. The full likelihood for
the model is then: \begin{equation}
L(Y_d | \theta_d) = p(\lambda(\mathbf{s},t)) \prod_{d=1}^{M}Pr(Y_d | \lambda(\mathbf{s},t), \theta_d).
\label{eq:FullLhood}
\end{equation} 

The process model is $p(\lambda(\mathbf{s},t))$: clearly there can only be one model at this level.  This formulation is clearly general enough to cover a
much wider range of models, so here, we will focus on the types of data
typical for species distributions, using a point process framework.

\subsection{Process Model}

Our process model follows \citet{Aarts2012, Dorazio2014} by using a
point process model. We assume that the locations of individuals are
points, possibly with marks (such as which species the individual is),
and we can model the distribution of points as a field with an
intensity, \(\lambda(\mathbf{s})\). For convenience in presenting the
process model, we describe this intensity with only spatial variation.
We model this field as a log-Gaussian Cox process with intensity
\(\lambda(\mathbf{s})=e^{\eta(\mathbf{s})}\)
\citep{moller2003statistical} with:

\begin{equation}
\eta(\mathbf{s}) = \sum_{i=1}^P \beta_i X_i(\mathbf{s}) + u(\mathbf{s}),
\label{eq:LCGP}
\end{equation} where \(X_i(\mathbf{s})\) is the field for the \(i^{th}\)
covariate
\citep[or, if more complex responses are needed, a feature \textit{sensu}][]{Elith2011}.

Residual spatial effects are modeled through \(u(\mathbf{s})\). This is
a random field which is set up so that there is a covariance between
points that depend on the distance between them (i.e.~there is a spatial
autocovariance). Although there are several alternatives to modelling this spatial effect, in practice, a
Gaussian Markov Random Field \citep{Lindgren2011} is a common and computationally efficient choice. Technically, this
means assuming \(u(\mathbf{s})\) is a Gaussian field with a Matèrn
correlation function: \begin{equation}
Cor(u(s_i), u(s_j))  = \frac{2^{1-\nu}}{\Gamma(\nu)} (\kappa \| s_i - s_j \|)^{\nu} K_{\nu} (\kappa \| s_i - s_j \|).
\label{eq:matern}
\end{equation}
with covariance function
\(Cov(u(s_i), u(s_j)) = \sigma_u Cor(u(s_i), u(s_j))\), where
\(\sigma_u^2\) is the marginal variance. In the analyses below we fix
\(\nu=1\), and parameterise the covariance as
\(\theta = \{ log(\tau), log(\kappa) \}\), where \(\tau\) is a local
variance parameter, so that \begin{equation}
\sigma_u^2 = \frac{1}{4 \pi \tau^2 \kappa^2},
\label{eq:maternMargVar}
\end{equation} and thus
\(\log(\tau)=-\log(4 \pi \sigma_u^2 \kappa^2)/2\). Finally, $u(\mathbf{s})$ is approximated through basis functions with local support and defined over a triangulation of the space. These basis functions are weighted by Gaussian random variables. The local nature of the basis functions guarantees the sparseness of the representation \citep{blangiardo2015spatial}.

Given this structure, the number of individuals in an area \(B\) follows
a Poisson distribution with mean: \begin{equation}
\mu(B) = \int_B \lambda(\mathbf{s}) ds = \int_B e^{\sum_{i=1}^P \beta_i X_i(\mathbf{s}) + u(\mathbf{s})} d\mathbf{s}. \label{eq:LCGPmean}
\end{equation} This then means that the probability that area \(B\) is
occupied is: \begin{equation}
Pr(N_B>0) = 1 - Pr(N_B=0) = 1-e^{-\mu(B)}.
\label{eq:PrOcc}
\end{equation} The integral in equation \eqref{eq:LCGPmean} is hard to handle analytically, so it is approximated by numerical integration in practice.
We use the approach of \cite{Simpsonetal2016OffTheGrid}, which makes use of the discretised version of the study region to approximate $\mu(B)$ in Equation \eqref{eq:LCGPmean}
as \begin{equation}
\mu(B) \approx \sum_{s=1}^{n_B} A(s) e^{\eta(B(s))}, \label{eq:LCGPmeanapprox}
\end{equation} where \(n_B\) is the number of integration points in
\(B\), each located at \(B(s)\), and \(A(s)\) is the area of the polygon
around \(s\), defined through a dual mesh. We thus only need to estimate the value of the intensity
at the integration points. If we want to estimate it for any other
point, we do it as an interpolation between the three points that form the
corners of the triangle that contains the point.

\subsection{Observation Process}

With the definition of the model for the actual abundance in any area,
we can add a variety of observation process models, appropriate to the
data at hand, fit the model to the data and estimate this abundance
layer.

\subsubsection{Point Counts}
\label{Sec:ObsPtCts}

Data is often collected in the form of counts of individuals of a
species, for example, the Breeding Bird Survey in North America
\citep{PardieckZiolkowski2015}. If we assume that observers observe at a
site \(\mathbf{s}\) for a time \(t\), and the probability that they
observe each individual in the site over the period is \(p\), then the
number of individuals counted follows a Poisson distribution. That is;
\begin{equation}
Pr(N(\mathbf{s},t)=r) = \frac{\eta(\mathbf{s},t)^r e^{-\eta(\mathbf{s},t)}}{r!}, \label{eq:AbundLhood}
\end{equation} where \begin{equation}
\eta(\mathbf{s},t) = pt \int_\mathbf{s} \lambda(\mathbf{s}) d\mathbf{s}. \label{eq:Abundmean}
\end{equation}

If we assume that the site is small, then we can treat the intensity and
covariates as constant over the whole site, so: \begin{equation}
\eta(\mathbf{s},t) \approx ptA(\mathbf{s}) \bar{\lambda}(\mathbf{s}). \label{eq:Abundmeanapprox}
\end{equation} where \(A(\mathbf{s})\) is the area of \(\mathbf{s}\). In
practice, we are unlikely to know \(A(\mathbf{s})\) or \(p\). But both
(as well as \(t\)) are part of the observation model, and can be seen as
different elements of the observation effort, so we collapse these into
a single parameter, \(E(\mathbf{s}) = ptA(\mathbf{s})\). Hence, we only have
a single parameter to be estimated. With repeated counts at a site, we
can estimate how these vary over time, introducing extra variance as an
overdispersion parameter, i.e.~\begin{equation}
\theta(\mathbf{s},t) \approx ptA(\mathbf{s}) \bar{\lambda}(\mathbf{s}) \epsilon(\mathbf{s}) \label{eq:AbundmeanapproxOD}
\end{equation} where \(\varepsilon(\mathbf{s})\) follows some
distribution, e.g.~a Gamma distribution (or equivalently, a \(\chi^2\)
distribution) leads to a negative binomial.

In practice, equation \eqref{eq:AbundmeanapproxOD} is modeled on the log scale: \begin{equation}
\log{\theta(\mathbf{s},t)} \approx \log(E) + \bar{\eta}(\mathbf{s}) +  \varepsilon(\mathbf{s}) ,
\label{eq:AbundmeanapproxODlink}
\end{equation} where \(\varepsilon(\mathbf{s})\) could be modeled as a
Gaussian noise term. If \(E\) varies between sites in a known way, this
can be added to the model. For example, if the observation time, \(t\),
varies, this can be included as an offset,
i.e.~\(\log(E) = \alpha + \log(t)\).

\subsubsection{Occupancy Models}
\label{Sec:ObsProcOcc}

Data from surveys of small sites may not be of the form of a count, but
simply of presence/absence. With a single visit to a site, the
likelihood is simply, from equation \eqref{eq:AbundLhood},
\(Pr(N(\mathbf{s},t)>0) = 1-Pr(N(\mathbf{s},t)=0) = 1 - e^{\eta(\mathbf{s},t)}\).
Working on the log scale for \(\eta(\mathbf{s},t)\), this becomes:
\begin{equation}
Pr(N(\mathbf{s},t)>0) = p(\mathbf{s},t) = 1-e^{-e^\eta(\mathbf{s},t))},
\label{eq:PrObscloglog}
\end{equation} which is the inverse of the cloglog link function
\citep[e.g.][\S 1.2.4]{McCullaghNelder1989}, i.e.~\begin{equation}
\log(-\log( 1-Pr(N(\mathbf{s},t)>0))) = \eta(\mathbf{s},t).
\label{eq:cloglog}
\end{equation}

If we have multiple visits to the site, we can extend this from a
Bernoulli to Binomial model, i.e.~likelihood that the species is
observed \(n\) times in \(N\) visits is: \begin{equation}
Pr(n=r|N,p(\mathbf{s},t))  \propto p(\mathbf{s},t)^n (1-p^{N-n}),
\label{eq:lhOcc}
\end{equation} where \(p = Pr(N(\mathbf{s},t)>0)\), from equation
\eqref{eq:PrOcc}. Note that this differs from a classical occupancy
model \citep{MacKenzie2002}, as we assume that occupancy in the area can
change (although a full occupancy model could also be developed).

\subsubsection{Point observations and presence-only models}
\label{Sec:ObsPtObs}

Data such as from eBird \citep{Sullivan2014eBird}, where presences alone
are recorded, can be treated as a point process
\citep[e.g.][]{Warton2010}. Models for point process data can be reduced
to a generalized linear model \citep{moller2003statistical}, and in the
context of distribution models have been shown to be equivalent to the
model assumed by MaxEnt \citep{Renner2013, Fithian2012}. If all
individuals are observed, then the observation model is trivial: the
model is just the process model. But this is seldom the case, and
instead, we use a thinned point process model: if each individual is
observed with probability \(q(\mathbf{s})\), the intensity of
observation is \(\phi(\mathbf{s})=q(\mathbf{s}) \lambda(s)\). If we
observe \(M\) points, at locations \(\mathbf{s}_1,...,\mathbf{s}_M\),
the log-likelihood is:
\begin{equation}
l(\beta | {\bf s}) \propto \sum_{i=1}^M \phi(\mathbf{s}) - \int_A \phi(a) da - \ln(M!). \label{eq:IPPlhood}
\end{equation} \citep{moller2003statistical}.

The integral cannot usually be calculated, so instead, we resort to a
numerical approximation, summing the intensity over discrete plates.
These become weighted sums of quadrature points (the corners of the
planes). If the quadrature points are placed at points \(r \in R\), the
likelihood becomes: \begin{equation}
l(\beta | {\bf s}) \approx \sum_{i=1}^{M+R} w_i [ z_{w,i} \ln{\phi(\mathbf{s})} - \phi(\mathbf{s})]
\label{eq:IPPlhoodapprox}
\end{equation} where \(w_i\) is a quadrature weight, and
\(z_{w,i}=w_i^{-1}\) if the point is a data point or 0 otherwise. This
is, by inspection, a Poisson likelihood, albeit with a non-integer
response. We can thus use a standard GLM formulation, with a log link,
for the model \citep{Renneretal2015}. On the log scale, the model for
the intensity of the observations is
\(\eta(\mathbf{s}) + \log(q(\mathbf{s}))\), so observation bias can be
added as additive terms \citep[e.g.][]{Fithian2012}, and made a function
of covariates (including, potentially, a residual spatial term).

\subsubsection{Regional lists}
\label{Sec:ObsProcRL}

We can also combine equations \eqref{eq:Abundmean} and \eqref{eq:PrOcc}
to model list of species from larger regions (e.g.~nature reserves or
states). It would be reasonable to assume that the data is certain, so
the likelihood is simply equation \eqref{eq:PrOcc}. Unfortunately, this
does not easily fit within the model fitting approach used here. If the
area is small enough, it could be approximated by a constant surface. If
the area is larger, an alternative approach is needed, e.g.~section
\ref{Sec:ObsProcERM}.

\subsubsection{Expert Range Maps}
\label{Sec:ObsProcERM}

In principle, we can adapt the model used in section \ref{Sec:ObsProcRL}
for expert range maps, but the problem of variation within the area
cannot be ignored. We thus take another approach, including the range
map as a covariate in the process model (equation \ref{eq:LCGP}). The
simplest way to do this is to use it as a binary variable: in/out. But
the range map may be wrong at a local scale, for example, if the scale of
the range map is too coarse, or if the species has expanded beyond its
range since the map was drawn. Thus, we use a more flexible model:
\begin{equation}
\rho(s) = - \gamma \min(0, ||s-r||),
\label{eq:RangeEff}
\end{equation} and we can assume \(\gamma > 0\). This can be done by
placing informative priors on the parameters. With sufficient data, we
would expect to overcome the priors, but for more data-poor species (limited data on species), they
should influence the predicted distribution. This approach is similar to
that taken by \citet{Merow2017}, except that they assume a more complex
functional form, but also fix the effect of distance from range edge,
rather than estimating it. Thus, the cost of using a less flexible model
is offset (hopefully) by the advantage of allowing the range maps to be
less accurate. Of course, informative prior distributions can be used to
increase the importance of the range map.

\subsection{Putting it all together: Using the framework to model biodiversity observation error}

To provide a conceptual link of the point process model framework in
data integration, we describe an integrated model that accounts for
biodiversity observation errors. We assume that biodiversity data
contains observation errors caused by imperfect detection,
uneven sampling effort and reporting bias. These sources of observation errors are assumed to thin the actual intensity
\(\lambda(\mathbf{s})\)
\citep{Dorazio2014, sicacha2021spatial, adjei2023structural} in a
hierarchical way \citep{adjei2023structural}. An observer first samples a given location to collect data (often at the most accessible locations) and collects data at these sampled locations. However, it is possible that the observer may be unable to
detect all the species of interest, and in reporting the detected
species, choose to ignore some of the detected species. 

Let
\(b(\textbf{s})\) be the sampling probability, \(\psi(\textbf{s})\) be
the detection probability and \(\gamma(\textbf{s})\) be the reporting
probability. Following the data generating model defined in equation \eqref{eq:FullLhood}, the likelihood of the integrated model in this illustration is defined as:
\begin{equation}\label{DataGeneratingFramework}
    \begin{split}
        L(Y_d| b(\textbf{s}), \psi(\textbf{s}), \gamma(\textbf{s})) =  p(\lambda(\textbf{s})) \prod_{d=1}^M  Pr(Y_d| \lambda(\textbf{s}), b(\textbf{s}), \psi(\textbf{s}), \gamma(\textbf{s})).
    \end{split}
\end{equation}

None of the observation models described in section \ref{Sec:ObsPtCts}
to \ref{Sec:ObsProcERM} - fitted with one dataset - can provide unique
estimates of the model parameters in the equation
\eqref{DataGeneratingFramework}. However, it is possible to integrate data from multiple types and sampling protocols to ensure that the model parameters are identifiable. 

If the sources of biases are assumed to be conditionally
independent on the datasets available, then the model likelihood in equation \eqref{DataGeneratingFramework} now
becomes: \begin{equation}\label{DataGeneratingFrameworkSub}
    \begin{split}
        L(Y_d| b(\textbf{s}), \psi(\textbf{s}), \gamma(\textbf{s})) &=  p(\lambda(\textbf{s})) \prod_{d_1=1}^{M_1}  Pr(Y_{d_1}| \lambda(\textbf{s}), b(\textbf{s}))\\  &\prod_{d_2=1}^{M_2}  Pr(Y_{d_2}| \lambda(\textbf{s}), \psi(\textbf{s})) \prod_{d_3=1}^{M_3}  Pr(Y_{d_3}| \lambda(\textbf{s}), \gamma(\textbf{s})),
    \end{split}
\end{equation} 
where \(M_1\) is the number of datasets that inform the
model on the estimation of the sampling probability, \(M_2\) is the
number of datasets that inform the model on the estimation of the
detection probability and \(M_3\) is the number of datasets that inform
the model on the estimation of the reporting probability.

For illustration, we assume that we have available occupancy (\(Y_{d_1}\)), count data
(\(Y_{d_2}\)) and presence-only (\(Y_{d_3}\)) to integrate together. We
further, assume that the presence-only data provide information on the
sampling process of observers, the occupancy data provides information
on the detection process and count data provides information on the
reporting process. Following the definition of the observation models in
equations \eqref{eq:AbundLhood}, \eqref{eq:PrObscloglog} and
\eqref{eq:IPPlhood}, the log-likelihood becomes:
\begin{equation}\label{eq:loglikeIDM}
\begin{split}
l(b(\textbf{s}), \psi(\textbf{s}), \gamma(\textbf{s})|Y_{d_1}, Y_{d_2}, Y_{d_3}) &\propto \ln(p(\lambda(\textbf{s}))) + \sum_{i = 1}^{n_{d_1}} \ln \bigg(1- e^{\int_\textbf{s} \psi(\textbf{s}) \lambda(\textbf{s}) d\textbf{s}} \bigg) \\
 &+ \sum_{i = 1}^{n_{d_2}} \ln \bigg(1- e^{\int_\textbf{s} \gamma(\textbf{s}) \lambda(\textbf{s}) d\textbf{s}} \bigg) - \sum_{i = 1}^{n_{d_2}} e^{\int_\textbf{s} \gamma(\textbf{s}) \lambda(s) ds} - n_{d_2} \ln (r!)\\
& + \sum_{i = 1}^{n_{d_3}} b(\textbf{s})\lambda(\textbf{s}) - \int_{A} b(a)\lambda(a)da - \ln(n_{d_3}!),
\end{split}
\end{equation} where \(n_{d_1}\) is the number of occurrence data
points, \(n_{d_2}\) is the number of count data points and \(n_{d_3}\)
is the number of location points in the presence-only data. This
log-likelihood can either be maximised in the frequentist approach or
simulated from in the Bayesian framework. It must be stated again that
the integral in equation \eqref{eq:loglikeIDM} needs to be approximated
with numerical methods. Moreover, the information used to estimate the
underlying state \(\lambda(\textbf{s})\) comes from the three datasets
integrated together, and this sharing of information produces precise
estimates of \(\lambda(\textbf{s})\). This has been noted by
\citet{Dorazio2014, fletcher2016integrated, koshkina2017integrated} as
the advantage of data integration.

\section{Case study}

Now, we provide an application of the point process framework to fit
data on freshwater fish in Norwegian lakes, to illustrate how the model can be tailored to fit the data sources we have available, and the ecological questions we wish to answer.

In this case study, we consider citizen science observations of four fish species (pike, \textit{Esox lucius}; perch, \textit{Perca fluviatilis}; brown trout, \textit{Salmo trutta}, and Actic char, \textit{Salvelinus alpinus}) available through Global Biodiversity Information Facility (GBIF; \url{https://www.gbif.org/}). The citizen science observations are presence-only occurrence records. Additionally, we also have access to a presence/absence data on freshwater fish in Norway from the Fish Status Survey of Nordic Lakes \citep{data_survey}. 

\begin{figure}
    \centering
    \includegraphics[width = \textwidth]{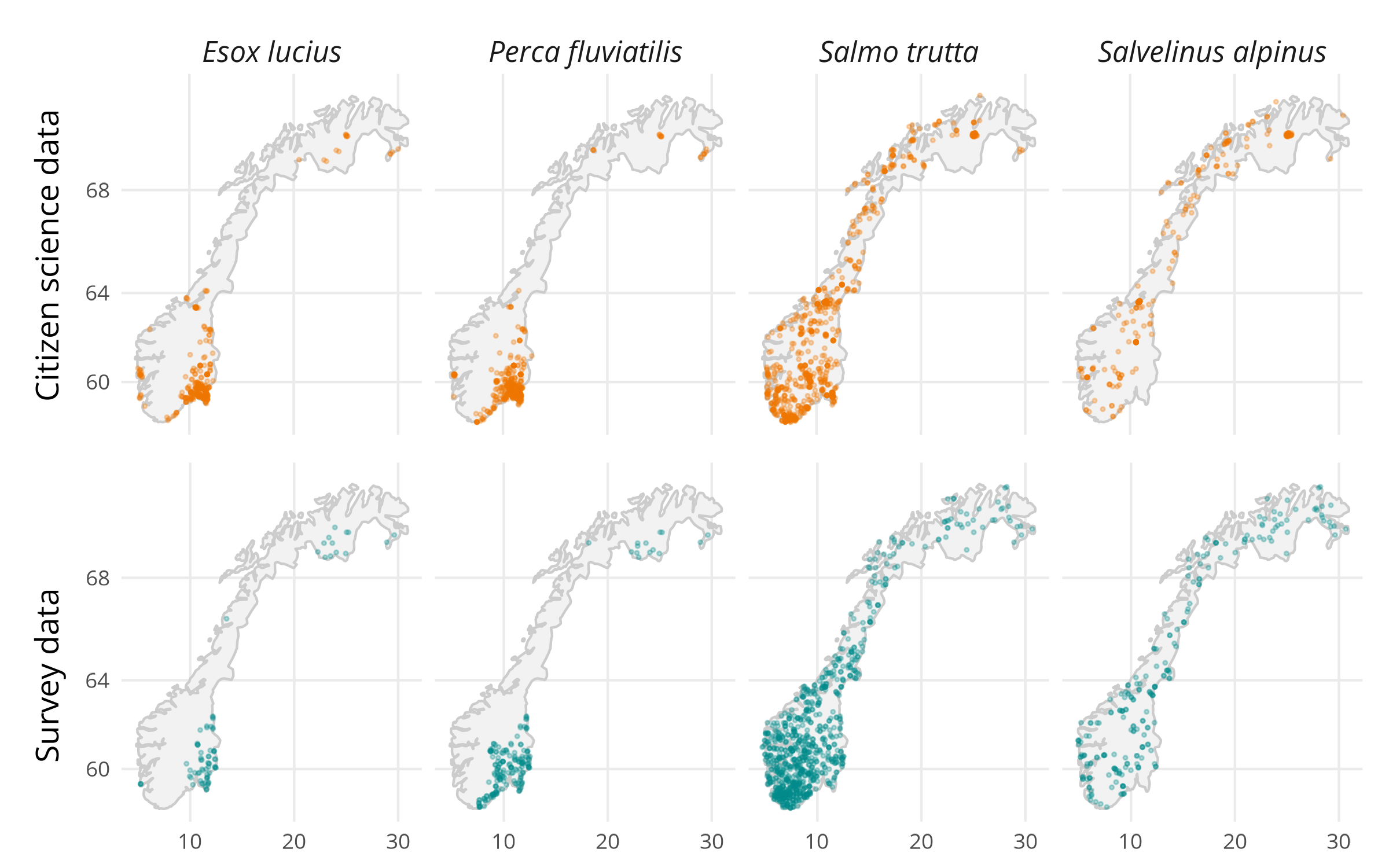}
    \caption{The observed presences of the different fish species in each of the two data sets. Locations are slightly jittered to avoid overplotting.}
    \label{fig:presence_points}
\end{figure}

The two data sources each have their advantages and disadvantages. 
The citizen science observations are opportunistic, and are likely to be spatially biased, since no sampling plan has been followed. However, we have a larger number of observations, particularly more recent observations in this dataset. The survey data, on the other hand, is from 1996, and so does not reflect the current distribution of these species. It does, however, present a less spatially biased sample than the citizen science observations. The presence points from both data sets can be seen in Figure \ref{fig:presence_points}. To take advantage of the strengths of both these data sets, we intend to integrate both in order to make predictions on the abundance of these freshwater fish in Norway.

For the presence/absence survey data, we use a Bernoulli distribution (as described in section \ref{Sec:ObsProcOcc}), where the presence probability depends on some covariates $x(s)$, along with a spatial field $\xi_{j}(s)$  for species $j\in \{\textit{Salmo trutta, Perca fluviatilis, Esox lucius, Salvelinus alpinus}\}$,
\begin{align}
    Y_{PA, j}(s_i) &\sim \text{Bernoulli}(p_{PA, j}(s_i)) \\
    \text{cloglog}(p_{PA, j}(s_i)) &= \alpha_{PA, j} + x(s_i)^T\beta_j + \xi_{j}(s_i).
\end{align}
The presence-only data is fitted with a Poisson point process model (as described in section \ref{Sec:ObsPtObs}), where the intensity depends on the same covariates $x(s)$ and the same spatial field $\xi_{j}(s)$, plus an additional spatial field $\xi_{\text{bias}}(s)$ that is unique to the citizen science data, but shared across all fish species:
\begin{align}
    Y_{PO, j}(s_i) &\sim \text{Poisson}(e^{\eta_{PO, j}(s_i)}) \\
    \eta_{PO, j}(s_i) &= \alpha_{PO, j} + x(s)^T\beta_j + \xi_{j}(s_i) + \xi_{\text{bias}}(s_i).
\end{align}

In this integrated distribution model, $\xi_{j}(s)$ describes the spatial autocorrelation shared between the datasets. $\xi_{\text{bias}}(s)$ is a bias field that captures the extra spatial variation in the presence-only records, and therefore captures the variation that is likely due to human observation biases rather than the distribution of the fish. The model was fitted using the \texttt{PointedSDMs} \citep{mostert2023pointedsdms} R package. Code and data for fitting the model, along with a more detailed description, can be found in the Supporting Information at \url{https://github.com/emmaSkarstein/IntegratedLakefish/}.

\begin{figure}
    \centering
    \includegraphics[width = \textwidth]{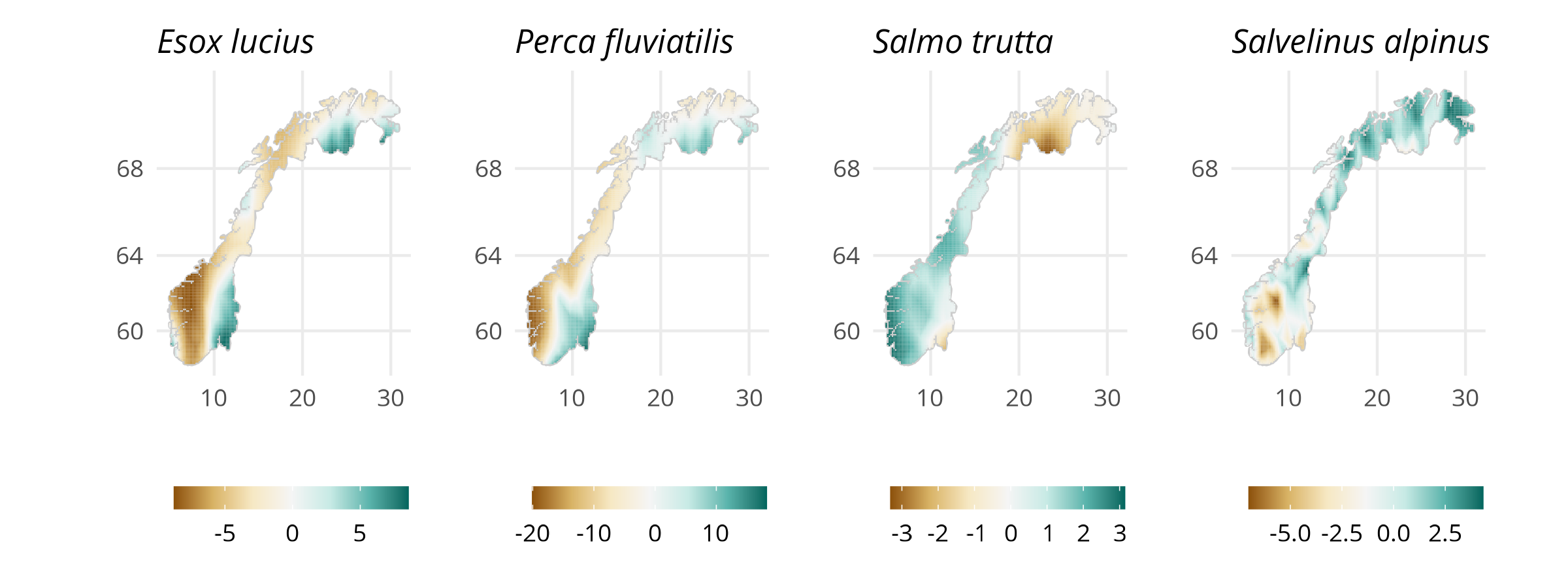}
    \caption{The mean log intensity of the estimated spatial fields for each of the fish species.}
    \label{fig:fishplot}
\end{figure}

Comparing the observations in the two data sets shown in figure \ref{fig:presence_points}, we see that the citizen science data shows certain geographical presences that the survey data does not. That is to be expected, since the survey data is from 1996, and there is reason to believe that the citizen science data shows a more updated view of the species distributions. We present the predictions of the mean log intensity fields for each species from the model in Figure \ref{fig:fishplot}.


\section{Discussion}

We present a statistical formulation of data integration using a point process framework, including a review of the current literature on data integration, an overview of the data types that are commonly integrated together, and a detailed description of the process model using the point process formulation and various observation models. As an illustration, we have applied the framework to predict the distributions of four fish species in Norwegian lakes, illustrating how this framework allows us to smoothly incorporate information about the observation processes for each of the data sources in a way that respects the data collection, and also makes sense ecologically.
In this paper, we focus on the point-process methodology estimated via a joint-likelihood approach, though other methods to construct ISDMs are also possible. Methods such as data pooling and combining independent models may fall within the definition of data integration, but are not recommended for most case scenarios \citep{fletcher2019practical}. While other popular methods may be justified, including using an informed prior in a Bayesian setup, using one dataset as a covariate for the other, or by connecting datasets together through a shared covariance matrix, designed to capture patterns present throughout the datasets \citep{pacifici2017integrating}. Simulation studies comparing the strengths and weaknesses of each of these methods have been completed \citep{pacifici2017integrating, ahmad2021integrated}; they found that the joint-likelihood method worked well when spatial bias in the presence only data was low or statistical components to account the bias were included in the model. 

Integrated distribution models fitted with the point process framework has emerged as an ideal approach to utilising the strengths of various data types to estimate state variables in a single analysis better \citep{Dorazio2014, fletcher2016integrated, koshkina2017integrated}. Another interesting feature of data integration is the adjustment of the biodiversity biases by using other data types. Data from well designed surveys would be an ideal complement to unstructured data \citep{koshkina2017integrated, fletcher2016integrated, pacifici2017integrating,giraud2016capitalizing}. Although data from structured surveys are less numerous, a good design will not be spatially biased, and this can be used to estimate and correct for the bias in the unstructured data \citep{simmonds2020more}. Thus, even in this model-based approach to modelling distributions, a design-based approach will be important if we are to utilise the data that is being collected fully. 

Any state variable of interest can be seen as derived quantities of point patterns \citep{kery2015applied}. With this, various data types can be integrated together through their basic derived quantity (point patterns) through the point process framework. In contrast to other papers summarised in Table \ref{tab:ISDM_table references} that focus on making inferences and predictions with integrated models, we focused on the statistical properties of the data in the integrated model. Here, we have described the data in terms of its statistical properties, but a more practically useful typology would describe them in terms of the methods used to collect it. Although we may see line transect data and eDNA data both as counts, there are of course large differences in how they are collected, and thus how we should treat them. For example, the correlation between read number (i.e.~count) of a sequence in eDNA data and abundance of the species is an area of active research \citep[e.g.][]{SkeltoneDNA}. A lot of work will therefore be needed to develop specific models for different data sources.

Our framework will only be helpful if it can be implemented in software that can be used by analysts, who may often be ecologists, potentially without formal training in statistics. The approach we have taken is closely linked to models that can be fitted with \texttt{R-INLA} \citep{inla} or \texttt{inlabru} \cite{inlabru}, but for the case study we have used the R-package \texttt{PointedSDMs} \citep{mostert2023pointedsdms}, which builds on \texttt{inlabru} but makes the model fitting a bit more convenient and user friendly. Additionally, the full potential of ISDMs will only be met if we can build a large-scale workflow that designs pipelines to move data from species occurrence databases (notably GBIF) to the modelling framework. An early implementation of such a workflow has been discussed and implemented through the R-package \texttt{intSDM} \citep{mostert2022intsdm}, which has been built to obtain species' occurrence and environmental data from popular repositories, standardise the data into a coherent framework, and obtain estimates from the model. 

We have outlined a framework for integrated distribution models and shown that it can be used for real problems. Within this framework, there is a lot of flexibility, and thus many developments to be explored. One of the key issues is how different datasets may support each other: surveys (for example) will still be important, because they provide reliable data that citizen science data can be calibrated to. But will this mean we should re-evaluate how surveys should be designed? This is the sort of issue that we will have to grapple with if we want to better understand and monitor biodiversity in the future.

\section*{Data and code availability}
Instructions for accessing the data and code used in the case study are available in the Supporting Information at \url{https://github.com/emmaSkarstein/IntegratedLakefish/}. All data is publicly available through GBIF.

\nocite{*}
\bibliographystyle{rss}
\bibliography{references}

\end{document}